\documentclass[sigconf]{acmart}
\AtBeginDocument{%
  }

\renewcommand\footnotetextcopyrightpermission[1]{}

\acmConference[CHI '26]{Proceedings of the 2026 CHI Conference on Human Factors in Computing Systems}{April 13--17, 2026}{Barcelona, Spain}

\usepackage{multirow}
\usepackage[utf8]{inputenc}
\usepackage{textgreek}
\usepackage{soul}
\usepackage{xcolor}
\usepackage{placeins}

\title{Seeing, Hearing, and Knowing Together: Multimodal Strategies in Deepfake Videos Detection}

\author{Chen Chen}
\affiliation{%
  \institution{Nanyang Technological University}
  \city{Singapore}
  \country{Singapore}}
\email{CHEN033@e.ntu.edu.sg}

\author{Dion Hoe-Lian Goh}
\affiliation{%
  \institution{Nanyang Technological University}
  \city{Singapore}
  \country{Singapore}
}
\email{ASHLGoh@ntu.edu.sg}

\begin{document}

\begin{abstract}
As deepfake videos become increasingly difficult for people to recognise, understanding the strategies humans use is key to designing effective media literacy interventions. We conducted a study with 195 participants between the ages of 21 and 40, who judged real and deepfake videos, rated their confidence, and reported the cues they relied on across visual, audio, and knowledge strategies. Participants were more accurate with real videos than with deepfakes and showed lower expected calibration error for real content. Through association rule mining, we identified cue combinations that shaped performance. Visual appearance, vocal, and intuition often co-occurred for successful identifications, which highlights the importance of multimodal approaches in human detection. Our findings show which cues help or hinder detection and suggest directions for designing media literacy tools that guide effective cue use. Building on these insights can help people improve their identification skills and become more resilient to deceptive digital media.
\end{abstract}



\maketitle
\thispagestyle{plain}
\begingroup
\renewcommand\thefootnote{}  
\footnotetext{This is a preprint of the paper accepted at CHI 2026. The final version will be available in the ACM Digital Library.}
\endgroup

\section{Introduction}
Deepfake videos refer to highly realistic, AI-generated videos that alter or swap human faces, which have quickly become a challenging form of online disinformation ~\cite{doss2023deepfakes, mirsky2021creation}. As these videos spread through social media, they can mislead audiences, damage people’s reputations, and undermine trust in digital content ~\cite{chesney2019deep, maddocks2020deepfake, vaccari2020deepfakes}. While public awareness of deepfakes is growing, people still struggle to recognise them ~\cite{goh2024he, kobis2021fooled}. Studies show that humans often mistake deepfakes for authentic videos ~\cite{nightingale2022ai, kobis2021fooled}, and many feel more confident in their judgments than their actual accuracy justifies ~\cite{kobis2021fooled}. This combination of vulnerability and overconfidence suggests that human judgment is often not able to separate real from fake in today’s digital media environment.

To address these challenges, scholars have proposed interventions to improve people's ability to detect deepfakes. For example, exposure to manipulated content before evaluation has been found to improve sensitivity ~\cite{tahir2021seeing}. Others have explored human-AI collaboration, showing that providing AI predictions ~\cite{lai2019human} or explanations ~\cite{lai2019human, groh2024human} can help people detect deepfakes more accurately. While these approaches demonstrate potential, they often focus on prescribing specific cues or providing guidelines for participants, without first establishing how people naturally approach the task and which strategies actually matter for performance.

Understanding the identification strategies people use is therefore a critical step. Prior work has shown that humans and AI detectors approach the task differently. Computer vision models can identify subtle pixel-level artefacts or irregularities that are hard to capture by the human eye ~\cite{mirsky2021creation}, while people rely on holistic impressions and contextual knowledge ~\cite{groh2022deepfake}. These differences highlight that humans and machines bring complementary strengths to the problem, and that for humans in particular, detection may depend on combining multiple strategies, such as noticing what they see (visual strategy), what they hear (audio strategy), and what they know (knowledge strategy) ~\cite{tang2024understanding, goh2024he}. Cues are associated with each strategy. Visual strategy involves cues like irregular blinking, unnatural lighting, or mismatched lip movements. The audio strategy draws on cues on the voice track, such as tone, fluency, or synchronization. Knowledge strategy depends on contextual cues such as knowing what a person is likely to say, recalling prior exposure to the video, or using fact-checking to cross-reference claims ~\cite{tang2024understanding, goh2024he}. These categories capture the main perceptual and external resources available to human judges, which together represent the core ways in which people process and evaluate media content.

Although prior work has identified several categories of strategies that people use to judge deepfakes, these insights are largely qualitative. We still lack a clear, quantitative understanding of how people employ these strategies in practice when evaluating deepfake videos. Moreover, past studies examined strategies in isolation, making it difficult to determine which cues matter most, how multiple strategies interact, and which ones become effective or misleading when evaluating real versus deepfake videos. Understanding these processes is crucial not only for theory advancement but also for designing effective media literacy interventions that can guide users toward more reliable cue use.

Past research shows that younger individuals frequently engage with online media and social networking platforms, making them particularly relevant subjects for studies on disinformation detection and media literacy ~\cite{perez2021fake}. We thus conducted a study with 195 young adults aged from 21 to 40 who judged real and deepfake videos, reported their confidence, and described the cues that shaped their decisions. Our analysis focuses on addressing three research questions:

RQ1: How do people perform when judging real versus deepfake videos, and how well is their confidence calibrated?

RQ2: How do different detection strategies (visual, audio, and knowledge), individually and in combination, contribute to accuracy and calibration in distinguishing real and deepfake videos?

RQ3: How do specific cues contribute to detection performance, and how do these cues interact when judging real versus deepfake videos?

By examining these questions, we highlight the importance of "seeing, hearing, and knowing together." Our findings reveal that no single cue guarantees success. Instead, integrating across cues strengthens detection.

This paper makes three contributions to the research on media literacy and disinformation. First, we offer empirical evidence on how people judge real and fake videos, including their accuracy and confidence calibration. Second, we show how visual, audio, and knowledge strategies and their combinations relate to performance in a quantitative way. Third, by examining cues within strategies and mapping their interactions within networks, we discover how specific cues support or hinder detection across real and fake videos. This analysis highlights which cues are diagnostic and which are misleading, and how their interplay affects performance. Finally, design implications are provided for media literacy interventions.

\section{Related work}
\subsection{Human Deepfake Detection Performance}
Prior work shows that humans can sometimes outperform AI detection models in specific contexts. For instance, when judging videos of celebrities ~\cite{groh2022deepfake}, interpreting social interactions ~\cite{das2022survey}, or evaluating individuals from familiar cultural or gender backgrounds ~\cite{lovato2024diverse}. At the same time, without structured training, overall detection accuracy are often around near chance levels ~\cite{diel2024human}, and people tend to be overconfident in their judgments, which increases vulnerability to manipulation ~\cite{diel2024human, kaur2024deepfake}. Confidence does not always align with accuracy, as individuals may confidently label a deepfake as real despite being wrong ~\cite{kobis2021fooled, bray2023testing}. The problem of overconfidence exacerbates susceptibility to false information. For example, overconfident individuals visit untrustworthy websites more often and have greater difficulty distinguishing between true and false claims ~\cite{lyons2021overconfidence}. They are also less likely to verify information with external sources ~\cite{olcaysoy2023updating}. 

On the other hand, some research finds a modest positive link between confidence and detection accuracy ~\cite{somoray2023providing}. From a cognitive psychology perspective, what matters is calibration, the alignment between confidence and actual correctness. Well-calibrated judgments are associated with higher decision quality ~\cite{basol2020good, saleh2024active}. Interventions such as game-based training can improve both confidence and calibration, helping people resist disinformation more effectively ~\cite{basol2020good, salovich2021misinformed}. Confidence thus functions both as a driver and outcome of disinformation processing, shaping how people engage with false content and being reshaped by their past experiences ~\cite{rapp2024confidence}.

\subsection{Use of Multiple Strategies in Human Detection of Deepfakes}
To strengthen both accuracy and calibration, it is essential to understand the factors that shape human decision-making in authenticity judgments. Prior research indicates that people rely on three main types of strategies when evaluating whether media is real or fake: visual, audio, and knowledge strategy ~\cite{tang2024understanding, perez2021fake, goh2024he}. Visual strategies include noticing artefacts such as unnatural blinking, distorted facial features, lighting inconsistencies, or overly smooth skin textures ~\cite{goh2024he}. Audio cues provide another important source of information. People may detect unusual intonation, odd pauses, or voices that sound slightly artificial ~\cite{warren2024better}. Misalignment between lip movements and speech, such as when words do not match the speaker's mouth shapes, can further raise suspicion ~\cite{somoray2025human}. Finally, knowledge-based strategy allows humans to use prior information about a person or event. For example, if a video shows a public figure making an implausible or out-of-character statement, they may doubt its authenticity ~\cite{chen2025understanding}.

These findings show that human judgment often combines perceptual and semantic cues, rather than relying on a single source of information. However, prior research has mostly examined these strategies qualitatively in isolation~\cite{tang2024understanding, goh2024he}. We know far less about how people combine multiple strategies in practice, and how such combinations affect both accuracy and confidence. That gap is especially important given the rise of multimodal deepfake detection research, where algorithms that integrate audio-visual signals consistently outperform those relying on single modalities ~\cite{raza2023multimodaltrace, oorloff2024avff}.

\section{Method}
\subsection{Participants}
We recruited 200 participants using convenience and snowball sampling. Recruitment was conducted through an online poster circulated on social media platforms. Each participant received SGD \$10 as reimbursement for completing the study. After removing incomplete or invalid responses, the final dataset included 195 participants. Each participant evaluated four videos, resulting in a total of 780 valid video judgments, along with confidence ratings and self-reported strategy use. The study was approved by the university's institutional ethics board.

Of the 195 participants, 111 were females (57\%) and 84 as males (43\%). The majority were between 21 and 30 years old (171 participants, 87.7\%), with the remaining 24 participants (12.3\%) aged 31 to 40. In terms of education, just over half held a bachelor's degree (104, 53.3\%), followed by those with junior college or equivalent qualifications (75, 38.5\%) and a smaller group with a master's degree (14, 7.2\%). A small number of participants reported other levels of education (2, 1.0\%). Table~\ref{tab:demographics} provides a detailed breakdown of participant demographics.

\begin{table}[ht]
\centering
\Description{Participants' demographic information, including gender, age group, and educational level.}
\caption{Participant demographics (N = 195).}
\label{tab:demographics}
\resizebox{\columnwidth}{!}{
\begin{tabular}{llcc}
\toprule
\textbf{Demographic Variable} & \textbf{Category} & \textbf{Count} & \textbf{\%} \\
\midrule
\multirow{2}{*}{Gender} 
  & Male   & 84  & 43.0 \\
  & Female & 111 & 57.0 \\
\midrule
\multirow{2}{*}{Age} 
  & 21--30 years & 171 & 87.7 \\
  & 31--40 years & 24  & 12.3 \\
\midrule
\multirow{4}{*}{Educational Level} 
  & Junior College or equivalent & 75  & 38.5 \\
  & Bachelor’s degree            & 104 & 53.3 \\
  & Master’s degree              & 14  & 7.2  \\
  & Others                       & 2   & 1.0  \\
\bottomrule
\end{tabular}
}

\end{table}

\subsection{Stimuli}
Participants viewed videos containing both visual and audio components. Each clip was brief (typically under 60 seconds) and featured a public figure delivering spoken messages. Our stimulus set comprised 20 videos in total: 10 authentic videos and 10 deepfakes collected from publicly accessible sources such as YouTube. The deepfake stimuli were chosen to reflect manipulation styles commonly seen in disinformation, including face swapping, lip-syncing, and facial reenactment, and covered a range of topics such as politics and sports to maximise diversity. Appendix B contains information on video content, duration, and example screenshots from two authentic and two deepfake videos.

\subsection{Procedure}
After providing informed consent, participants completed the study individually through an online platform (A screenshot was provided in Appendix D). Each participant viewed four short videos randomly drawn from our stimulus set of 20 mentioned above. For every participant, two authentic and two deepfake videos were presented in random order to minimise order effects.
For each video, participants were asked to: (1) judge authenticity by indicating whether the video was real or fake, (2) rate their confidence on a five-point Likert scale (1 = not confident at all, 5 = very confident), and (3) identify the strategies that informed their judgment using a structured questionnaire. The questionnaire included three categories of cues: visual (e.g., facial features), audio (e.g., vocal tone), and knowledge-based (e.g., prior knowledge of the person or event). These cues were based on prior research on deepfake identification ~\cite{tang2024understanding, goh2024he}. Participants could select multiple cues within each strategy. This procedure was repeated four times per participant (once for each video). The complete cue items presented to participants are listed in Appendix C, and each item was shown as a checkbox option with a short description. After completing all video evaluations, participants also provided demographic information such as age, gender, educational background.

\subsection{Metrics}
\subsubsection{Expected Calibration Error (ECE)}
We evaluate calibration using the Expected Calibration Error (ECE) ~\cite{guo2017calibration, naeini2015obtaining}. 
ECE is computed by partitioning predictions into $M$ equally sized bins according to their reported confidence scores, and averaging the absolute difference between accuracy and confidence in each bin. Formally,

\begin{equation}
\mathrm{ECE} = \sum_{m=1}^{M} \frac{|B_m|}{N} \; \big| \; \mathrm{acc}(B_m) - \mathrm{conf}(B_m) \; \big| ,
\end{equation}

where $N$ is the total number of predictions, $B_m$ is the set of samples in the $m$th confidence bin, 
$\mathrm{acc}(B_m)$ is the empirical accuracy in that bin, and $\mathrm{conf}(B_m)$ is the average reported confidence.
A higher ECE indicates greater miscalibration, meaning that confidence levels are not aligned with empirical accuracy.

\subsubsection{Accuracy}
Alongside calibration, we also report \emph{Accuracy}, defined simply as the proportion of correctly identified items:

\begin{equation}
\mathrm{Accuracy} = \frac{1}{N} \sum_{i=1}^{N} \mathbf{1}\{\hat{y}_i = y_i\} ,
\end{equation}

where $\hat{y}_i$ is the predicted label and $y_i$ is the true label for the $i$th item, and $\mathbf{1}\{\cdot\}$ is the indicator function.
Accuracy provides a straightforward measure of overall detection performance, while ECE quantifies the alignment between subjective confidence and objective correctness.

\subsection{Data Analysis}
Our analysis proceeded in three steps. First, we compared overall performance between real and deepfake videos. Specifically, we examined differences in accuracy and ECE using paired-sample t-tests.

Second, we examined the role of self-reported strategies in shaping performance outcomes. To achieve this, we conducted analyses of variance (ANOVA) to compare accuracy across participants who relied on visual, audio, or knowledge-based strategy. Post-hoc tests were then applied to identify which specific strategies or combinations of strategies accounted for significant differences on performance.

Finally, to move beyond individual strategies, we applied association rule mining to identify frequent co-occurrences of cues of three strategies and their performance. In our analysis, the antecedents were the specific cues participants reported using, while the consequent was the detection outcome (correct or incorrect identification). To ensure meaningful patterns, we set thresholds of 0.20 for support and 0.60 for confidence. A support threshold of 0.20 means that only antecedent cue combinations present in at least 20\% of all trials are considered, which helps filter out rare cue sets that might lead to overfitting conclusions. The confidence threshold of 0.60 ensures that the rule predicts the detection outcome correctly in at least 60\% of those cases. These values were derived via pilot exploration of our dataset and align with common practice in the literature ~\cite{huang2000fast}. This yielded 66 association rules for deepfakes and 128 for real videos. We then used these antecedent sets to construct sub-datasets that included only trials where all cues in the antecedent were selected. For each set, we computed mean accuracy and ECE. To visualise inter-cue relationships, antecedent sets with multiple cues were decomposed into all possible two-cue pairs. These pairs were represented as edges in a network graph, with node size reflecting cue frequency and edge thickness reflecting average performance (accuracy or ECE). This approach enabled us to capture not only individual cue effects but also how cues interact in networks.

\section{Results}
\subsection{How Well Do People Detect Deepfakes? (RQ1)}

Table~\ref{tab:performance_results} summarises participants' performance across overall, real, and deepfake video conditions. Overall accuracy was $0.76$. Participants were significantly better at identifying real videos ($M = .79$, $SD = .31$) than detecting deepfakes ($M = .72$, $SD = .33$), $t(194) = -2.08$, $p = .038$.  

In terms of confidence, however, participants reported nearly identical levels when judging real videos ($M = 3.57$, $SD = .91$) and deepfakes ($M = 3.56$, $SD = .87$). The difference was not statistically significant, $t(194) = -0.14$, $p = .89$. This indicates that although accuracy dropped for deepfakes, participants felt just as confident as when judging real content.  

Calibration analyses highlighted this mismatch. ECE was higher for deepfake judgments ($M = .32$, $SD = .29$) than than for real videos ($M = .24$, $SD = .26$), $t(194) = 2.96$, $p = .003$. In other words, participants' confidence was less aligned with actual accuracy when evaluating deepfakes.

\begin{table*}[t]
\centering
\Description{Performance results for accuracy, confidence, and ECE of overall, deepfakes and real video conditions.}
\caption{Performance comparison between real and deepfake videos. Values are reported as participant-level mean $\pm$ SD. Statistical tests are paired-sample $t$-tests.}
\label{tab:performance_results}
\begin{tabular}{lcccc}
\toprule
\textbf{Measure} & \textbf{Real (M $\pm$ SD)} & \textbf{Deepfake (M $\pm$ SD)} & \textbf{$\Delta$ (Real–Fake)} & \textbf{$t$-test} \\
\midrule
\multicolumn{5}{l}{\textbf{Overall}} \\
Accuracy   & \multicolumn{2}{c}{0.76 $\pm$ 0.24} & — & — \\
Confidence & \multicolumn{2}{c}{3.57 $\pm$ 0.77} & — & — \\
ECE    & \multicolumn{2}{c}{0.27 $\pm$ 0.19} & — & — \\
\midrule
\multicolumn{5}{l}{\textbf{Real vs. Deepfake}} \\
Accuracy   & 0.79 $\pm$ 0.31 & 0.72 $\pm$ 0.33 & 0.07 & $t(194)=-2.08,\,p=.038$ \\
Confidence & 3.57 $\pm$ 0.91 & 3.56 $\pm$ 0.87 & 0.01 & $t(194)=-0.14,\,p=.89$ \\
ECE        & 0.24 $\pm$ 0.26 & 0.32 $\pm$ 0.29 & –0.08 & $t(194)=2.96,\,p=.003$ \\
\bottomrule
\end{tabular}
\end{table*}

\begin{figure}[ht]
  \centering
  \Description{Result of confusion matrix.}
  \includegraphics[width=\linewidth]{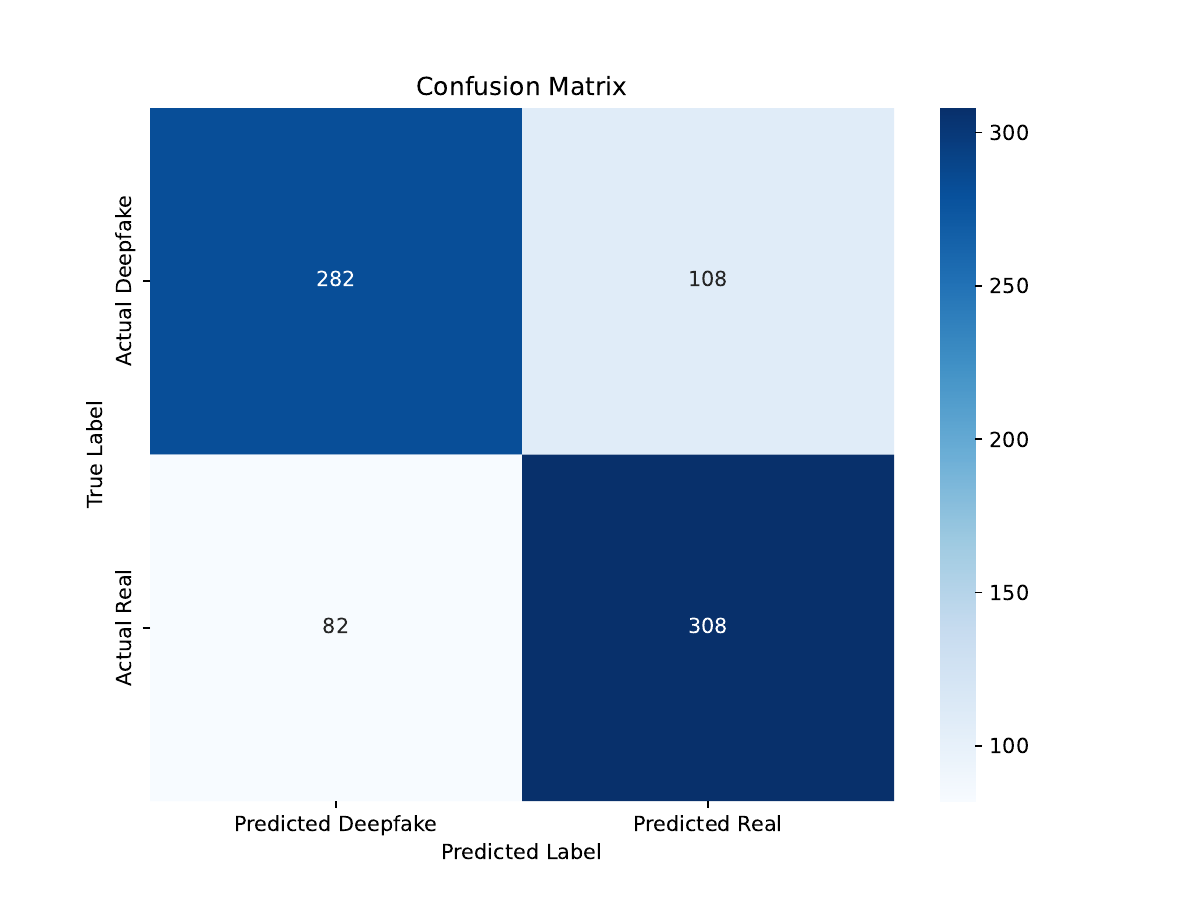}
  \caption{Confusion Matrix.}
  \label{fig:figurecon}
\end{figure}

Figure~\ref{fig:figurecon} shows the confusion matrix for all judgments. Participants correctly identified 282 deepfakes and 308 real videos, with 108 deepfakes missed and 82 real videos falsely labelled as fake. Participants were more likely to incorrectly judge a deepfake as real than to mistakenly classify real videos as fake.

\subsection{Do Multiple Strategies Improve Detection? (RQ2)}

Table~\ref{tab:strategy_real_fake} presents detection performance across the strategies that participants reported using. Visual only (V), audio only (A), and knowledge only (K) refer to participants selecting cues solely from that strategy. Next, visual + audio (V+A) indicates the selection of both visual and audio cues but no knowledge cues. Finally, visual + audio + knowledge (V+A+K) reflects the use of cues of all three strategy types. About 80\% of trials for deepfake videos involved the use of all three strategies simultaneously, while for real videos, this proportion was 75\%. 

For deepfake videos, visual-only and audio-only usage achieved the highest accuracy, though the sample size was too small for reliability. Participants who used both visual and knowledge strategies achieved the second-highest accuracy and low ECE (0.95, 0.18; $N = 39$), but again the sample size was limited. In contrast, the same strategy combination for real videos produced lower accuracy and higher ECE (0.50, 0.21; $N = 18$). Multiple strategies use (single, double, or triple) did not make a significant difference for deepfake detection. Accuracy was not affected by strategy grouping, $F(3, 386) = 1.96, p = .120$, and ECE also showed no significant group differences, $F(3, 241) = 1.42, p = .238$. These findings suggest that deepfakes remained consistently difficult to detect, even when participants used multiple strategies. 

For real videos, however, strategy differences played a more important role. A one-way ANOVA revealed a significant effect of strategy on accuracy, $F(3, 386) = 3.41, p = .018$. Multiple strategies approaches, especially those combining knowledge and audio strategy, visual and audio strategy, or all three strategies, showed higher performance compared with visual- or audio-only use, with accuracies approaching 0.90 in some cases. This suggests that cross-checking what participants saw or heard against knowledge source helped them recognise authentic content more effectively. By contrast, reliance on a visual-only strategy led to very poor performance for real videos (0.13 accuracy), although this subgroup was small ($N = 8$). 

The post-hoc tests (Table~\ref{tab:posthoc_results}) provide further statistical comparisons of group strategies. For real videos, the overall ANOVA indicated significant group differences in accuracy, yet none of the pairwise comparisons reached conventional significance. The closest effect was between Group~3 (triple-strategy) and Group~1 (single-strategy), where the difference approached significance ($p = .057$), suggesting a trend that applying multiple strategies may confer some advantage. In addition, Group~2 (double-strategy) produced significantly better calibration than Groups~1 and~3, $p = .038$, meaning their confidence was more closely aligned with actual accuracy. 

Taken together, the results for RQ2 show a clear asymmetry. While strategy combinations did not improve deepfake detection, they did matter for real videos. Multiple strategies, particularly those that included knowledge cues, were associated with stronger accuracy and better calibration.

\begin{table*}[t]
\centering
\Description{Detection performance by strategies and strategy group (single, double, triple strategies), including mean accuracy, mean ECE, and ANOVA results for real vs. deepfake videos.}
\caption{Accuracy and calibration error (ECE) by strategy type and combinations, split by deepfake vs.\ real videos. Values are trial-level means. One-way ANOVA tested differences across grouped strategies.}
\label{tab:strategy_real_fake}
\begin{tabular}{lcccccc}
\toprule
\multirow{2}{*}{\textbf{Strategy Use}} &
\multicolumn{3}{c}{\textbf{Deepfake}} &
\multicolumn{3}{c}{\textbf{Real}} \\
\cmidrule(lr){2-4}\cmidrule(lr){5-7}
& \textbf{Acc (M)} & \textbf{ECE} & \textbf{$N$} &
  \textbf{Acc (M)} & \textbf{ECE} & \textbf{$N$} \\
\midrule
None (no strategy)              & 0.33 & 0.89 &  3 & 0.50 & 0.27 & 6 \\
Visual only                     & 1.00 & 0.17 &  6 & 0.13 & 0.48 &  8 \\
Audio only                      & 1.00 & 0.13 &  2 & 0.67 & 0.22 &  9 \\
Knowledge only                  & 0.67 & 0.20 &  9 & 0.92 & 0.27 & 12 \\
Visual + Audio only             & 0.44 & 0.42 &  9 & 0.83 & 0.17 & 18 \\
Visual + Knowledge only         & 0.95 & 0.18 & 39 & 0.50 & 0.21 & 18 \\
Audio + Knowledge only          & 0.60 & 0.20 & 10 & 0.88 & 0.23 & 25 \\
\midrule
Group 1 (singles: A/V/K)        & 0.82 & 0.12 & 17 & 0.62 & 0.14 & 29 \\
Group 2 (doubles: V+A / V+K / A+K)    & 0.81 & 0.06 & 58 & 0.75 & 0.14 & 61 \\
Group 3  (V+A+K)               & 0.71 & 0.18 & 312 & 0.82 & 0.12 & 294 \\
\midrule
Overall                         & 0.72 & 0.14 & 390 & 0.79 & 0.13 & 390 \\
\midrule
\multicolumn{7}{l}{\textit{ANOVA (Deepfake videos):} Accuracy $F(3, 386)=1.96$, $p=.120$; ECE $F(3, 241)=1.42$, $p=.238$} \\
\multicolumn{7}{l}{\textit{ANOVA (Real videos):} Accuracy $F(3, 386)=3.41$, $p=.018$; ECE $F(3, 232)=2.43$, $p=.066$} \\
\bottomrule
\end{tabular}
\end{table*}

\begin{table*}[t]
\centering
\Description{Post-hoc Tukey HSD test results comparing strategy groups of real videos, showing mean differences and significance for accuracy and ECE.}
\caption{Tukey HSD post-hoc test results for strategy groups of real videos. 
Values show mean differences. Note: Other group comparisons were also tested, but all yielded $p$-values greater than $0.10$.}
\label{tab:posthoc_results}
\begin{tabular}{llccc}
\toprule
\textbf{Measure} & \textbf{Comparison} & \textbf{Mean Diff} & \textbf{$p$-value} & \textbf{Sig.} \\
\midrule
Accuracy     & Group 1 -- Group 3 & 0.199 & 0.057 & \textbf{Marginal} \\
ECE          & Group 2 -- Group 1 and Group 3  & 0.281 & 0.038 & \textbf{Weak} \\
\bottomrule
\end{tabular}
\end{table*}

\subsection{Which Cues Help or Mislead Humans? (RQ3)}

Table~\ref{tab:subcue_results_full} shows how specific visual, audio, and knowledge cues influenced participants' detection performance in terms of accuracy and ECE. More details on cue definitions are provided in Appendix~A.

\textbf{Visual cues.} Reliance on visual appearance (e.g., facial irregularities) yielded relatively strong results, with an accuracy of .84 for deepfakes and .78 for real videos. The difference was statistically significant, $t(194) = 2.10, p = .036$, suggesting that appearance cues were particularly helpful for spotting manipulated content compared to real videos. In contrast, visual environment cues (e.g., background inconsistencies) produced lower accuracy overall (.58 vs.\ .55), with no significant differences between video types. Visual production quality cues performed the worst overall, though accuracy was slightly higher for deepfakes (.48 vs.\ .41). However, reliance on production quality was associated with significantly higher calibration error, $t(194) = 2.19, p = .030$, indicating that participants who focused on production quality tended to be overconfident.

\textbf{Audio cues.} Vocal cues (e.g., tone) improved real-video accuracy (.81) compared to deepfakes (.69), $t(194) = -3.58, p < .001$, and also showed better calibration, $t(194) = 3.56, p < .001$. Similarly, language cues (e.g., phrasing) supported real-video accuracy (.55 vs.\ .47, $p = .026$) and improved calibration, $t(194) = 4.48, p < .001$. By contrast, sound-quality cues did not distinguish real from fake (.49 vs. .52, $t(194) = -0.86, p = .391$ ) but produced worse calibration for deepfakes, $t(194) = 3.83, p = .0002$. These findings suggest that while vocal and linguistic signals boosted accuracy, they also helped align confidence more closely with actual performance for real videos.

\textbf{Knowledge cues.} External knowledge sources (e.g., fact-checking) offered slightly better accuracy for deepfakes (.47 vs.\ .39) but no calibration benefits. Prior knowledge of people or events (.61 vs.\ .62, n.s.) was relatively stable across conditions and not significantly different. Intuition-based judgments, used frequently, achieved relatively strong accuracy (.70 for deepfakes, .59 for real), with a significant difference, $t(194) = 3.24, p = .001$. This suggests that instinct sometimes helped participants resist manipulation but could also bias their judgments of authentic videos.

\subsection{What Happens When Cues Combine? (RQ3)}

\begin{figure}[ht]
  \centering
  \Description{Bar charts comparing mean accuracy (top) and ECE (bottom) by antecedent sets for deepfake vs. real videos.}
  \includegraphics[width=\linewidth]{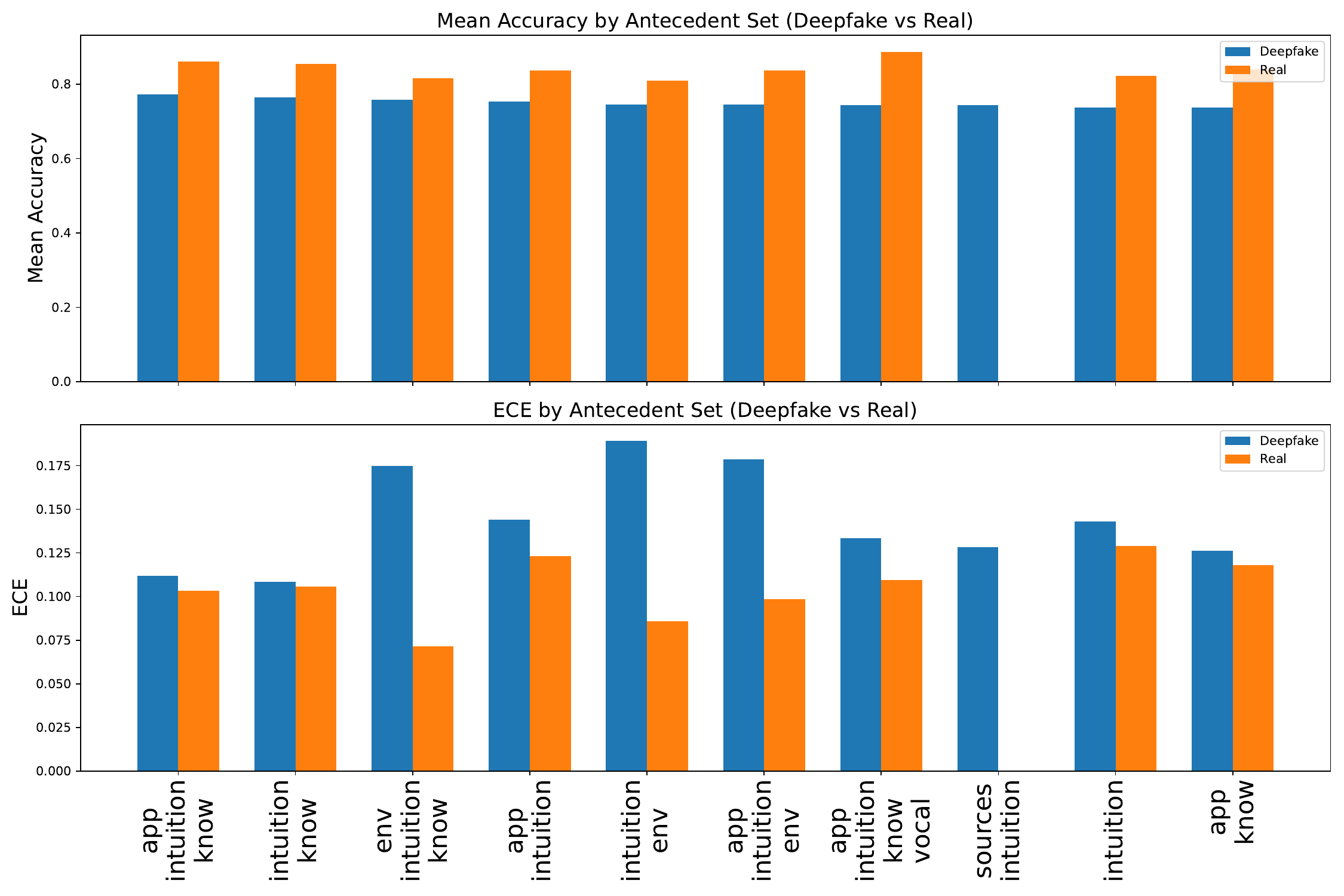}
  \caption{Top 10 accuracy by strategy cue combinations for deepfake vs. real videos. The chart compares how often specific multiple cue sets led to detection performance. Bars show mean accuracy and ECE for each cue combination, separately for real and deepfake videos.}
  \label{fig:figure2}
\end{figure}

Figure~\ref{fig:figure2} presents the top 10 strategy cue combinations ranked by mean accuracy and ECE, split by deepfake and real videos. The selection includes the top 10 combinations from each video type based on accuracy. The results show a consistent pattern where real videos achieve higher accuracy and lower ECE across most cue combinations compared to deepfakes. For example, sets combining prior knowledge, intuition, and environment cues or appearance, intuition, and environment cues demonstrate strong calibration on real videos, with accuracy values above 0.8 and ECE values notably lower than for deepfakes. While some combinations, such as (knowledge + intuition + appearance) or (intuition + knowledge), show smaller gaps between real and deepfake performance.

\begin{figure}[ht]
  \centering
  \Description{Two network graphs showing cue connections for deepfake (left) and real (right) videos. Node colour represents modality (visual, audio, knowledge), node size reflects cue frequency, and edge thickness reflects mean accuracy.}
  \includegraphics[width=\linewidth]{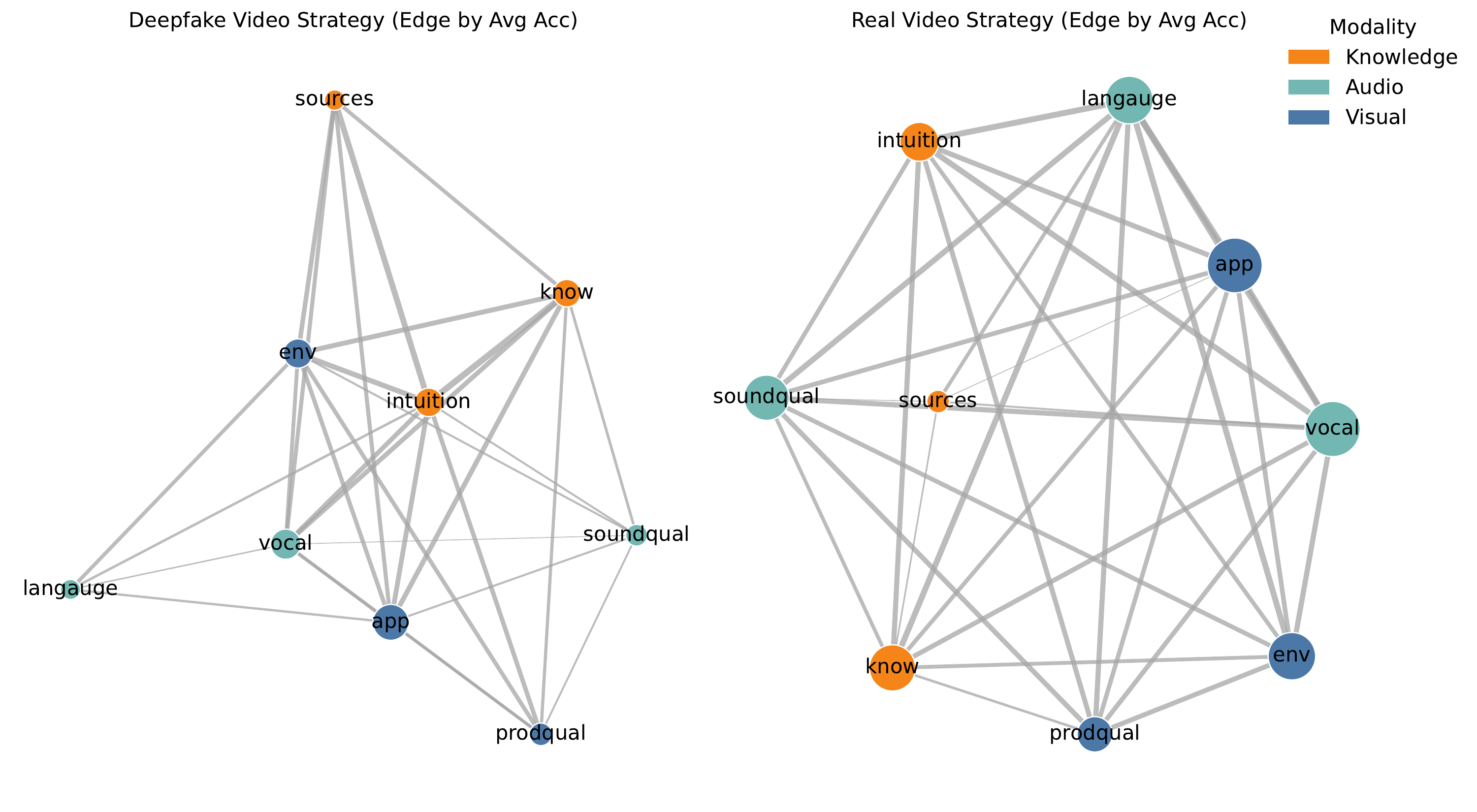}
  \caption{Network visualisation of strategy cue accuracy for deepfake videos and real videos. Node size represents cue frequency of use. Edge thickness represents average accuracy when cues were combined. Colours indicate modality (blue = visual, teal = audio, orange = knowledge). Node labels display cue names (app-appearance, env-environment, prodqual-production quality; soundqual-sound quality; sources-information sources, know-prior knowledge of person/video content/ video context). This figure highlights which cue combinations supported successful detection for each video type.}
  \label{fig:figure3}
\end{figure}

\begin{figure}[ht]
  \centering
  \Description{Two network graphs similar to Figure 2 but weighted by calibration (ECE inverted). Thicker edges indicate better alignment between confidence and actual accuracy.}
  \includegraphics[width=\linewidth]{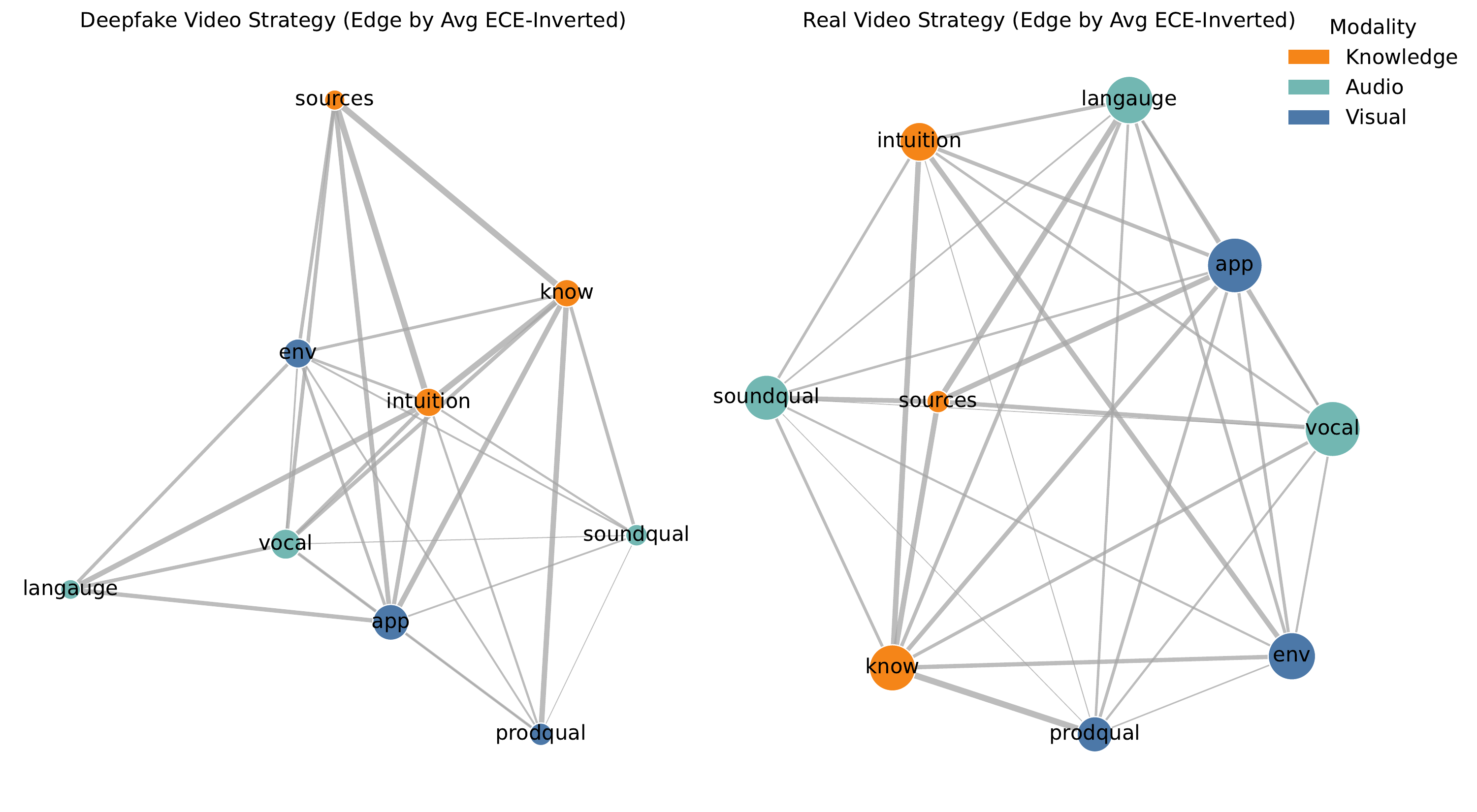}
  \caption{Network visualisation of strategy cue calibration (ECE-inverted) for deepfake videos and real videos. Node size represents cue frequency of use. Edge thickness represents better alignment between confidence and accuracy (lower ECE). Colours indicate modality (blue = visual, teal = audio, orange = knowledge). This figure illustrates which cues contributed to more calibrated judgments across video types.}
  \label{fig:figure4}
\end{figure}

Figure~\ref{fig:figure3} and figure ~\ref{fig:figure4} visualise the networks of strategy cues for deepfake videos (left) and real videos (right), showing their roles in accuracy and ECE.

For deepfake videos, a few local clusters emerge more prominently. For instance, connections among visual appearance and environment, prior knowledge, vocal, and intuition form stronger ties. This suggests that participants sometimes combined surface-level irregularities with contextual knowledge and intuitive judgments when evaluating deepfakes, and achieved higher accuracy. For real videos,  central hubs include more cues such as audio-vocal, audio-language, visual appearance, prior knowledge, and intuition.

The calibration network for deepfake videos (ECE inverted) shows that knowledge-based cues, such as intuition and sources, are linked with relatively thicker edges. This suggests that these cues supported better alignment between confidence and accuracy in some cases. By contrast, edges involving visual production quality, or sound quality remain thin. This indicates weaker calibration when participants relied on these signals. For real videos, thick edges are visible among audio vocal, audio language, visual appearance, and prior knowledge related cues. This shows robust calibration when participants integrated these cues.

\begin{table*}[t]
\centering
\Description{Detection performance of cues of three strategies.}
\caption{Accuracy, ECE, and frequency for strategy cues across deepfake vs.\ real videos. Values are trial-level means. Paired-sample $t$-tests compare deepfake and real conditions. Bold indicates $p < .05$}
\label{tab:subcue_results_full}
\begin{tabular}{lccccccccc}
\toprule
\multirow{2}{*}{\textbf{Strategy Cues}} & 
\multicolumn{3}{c}{\textbf{Deepfake}} & 
\multicolumn{3}{c}{\textbf{Real}} & 
\multicolumn{2}{c}{\textbf{$t$-test}} \\
\cmidrule(lr){2-4}\cmidrule(lr){5-7}\cmidrule(lr){8-9}
& Acc (M) & ECE & $N$ & Acc (M) & ECE & $N$ & Accuracy $t,p$ & ECE $t,p$ \\
\midrule
Visual appearance      & 0.84 & 0.14 & 327 & 0.78 & 0.12 & 304 & \textbf{2.10, .036} & \textit{1.83, .068} \\
Visual environment     & 0.58 & 0.18 & 228 & 0.55 & 0.09 & 214 & 1.01, .312 & 1.22, .225 \\
Visual production qual.& 0.48 & 0.19 & 189 & 0.41 & 0.12 & 158 & \textbf{2.24, .026} & \textbf{2.19, .030} \\
Audio vocal            & 0.69 & 0.18 & 271 & 0.81 & 0.15 & 314 & \textbf{-3.58, .0004} & \textbf{3.56, .0004} \\
Audio language         & 0.47 & 0.15 & 184 & 0.55 & 0.14 & 215 & \textbf{-2.22, .026} & \textbf{4.48, <.001} \\
Audio sound quality    & 0.49 & 0.19 & 191 & 0.52 & 0.13 & 203 & -0.86, .391 & \textbf{3.83, .0002} \\
Knowledge (sources)    & 0.47 & 0.12 & 185 & 0.40 & 0.11 & 155 & \textbf{2.17, .030} & 1.08, .281 \\
Knowledge (prior know.)& 0.61 & 0.13 & 237 & 0.62 & 0.13 & 243 & -0.44, .659 & \textbf{2.82, .005} \\
Knowledge (intuition)  & 0.70 & 0.14 & 274 & 0.59 & 0.13 & 231 & \textbf{3.24, .001} & \textbf{2.84, .005} \\
\bottomrule
\end{tabular}
\end{table*}

\section{Discussion}
\subsection{Asymmetries in Performance between Real and Deepfake Videos}

Our findings highlight both the promise and limits of human detection abilities. Participants achieved above-chance accuracy, consistent with prior studies showing that humans can sometimes be good at identifying manipulated media featuring celebrities ~\cite{groh2022deepfake}. However, detection was uneven across video types. Real videos were identified more reliably than deepfakes. At the same time, confidence ratings remained stable across video types. Past research also shows the overconfidence pattern in deepfake video detection ~\cite{kobis2021fooled}. Extending past work, our study used calibration metrics (ECE) rather than only correlations, showing that participants were poorer calibrated for deepfakes (higher ECE) but better calibrated for real videos (lower ECE).

Two reasons may explain this asymmetry. First, prior work documents a "realness bias" in which people assume authenticity unless strong artefacts are visible ~\cite{chesney2019deep}. Second, because our stimuli featured well-known figures, participants may have overestimated their detection ability when they recognised a familiar face. Past research provides evidence that confidence aligns more closely with accuracy when unfamiliar targets are used ~\cite{somoray2023providing}.

\subsection{Benefits and Constraints of Multiple Strategies Integration}
By analysing strategies separately and in combination, we show that human detection is inherently multimodal, as most participants reported using all three strategy types. This aligns with findings in cognitive psychology that judgments integrate perceptual and knowledge-based cues \cite{vaccari2020deepfakes}. 

Yet, effectiveness varied across strategy combinations. For deepfakes, no significant performance differences were observed between single, double, and triple strategy groups. In contrast, real video judgments benefitted from multimodal integration, where combinations of visual, audio, and knowledge strategies boosted accuracy and improved calibration. This difference may reflect the novelty and sophistication of deepfakes. A possible explanation for this asymmetry is that real videos often provide coherent cross-modal evidence, where visual appearance, vocal tone, language flow, and contextual cues naturally align. When these cues reinforce one another, integrating multiple cues strengthens the overall judgment signal and improves performance. In contrast, deepfakes may contain locally convincing elements such as realistic facial textures or fluent speech while still exhibiting subtle cross-modal inconsistencies that are too faint or unfamiliar for participants to interpret reliably~\cite{chen2025generation}. When cues conflict or feel ambiguous, adding more modalities may introduce additional noise rather than clarity. Moreover, this difference in the performance of using multiple strategies for deepfakes versus real videos may also reflect the novelty and sophistication of deepfakes. Participants could have been less familiar with advanced manipulation techniques, limiting the effectiveness of combining strategies. These findings highlight the importance of teaching users not just to notice different strategies but also to understand when and how to apply them effectively.

\subsection{Cue-Level Patterns}
Cue-level analyses further reveal modality-specific strengths and weaknesses. For deepfakes, visual appearance (e.g., facial irregularities) and intuition cues were relatively effective, with audio-vocal cues also contributing. This pattern is consistent with the fact that deepfake manipulations often target facial regions such as eyes or mouths, which can produce visible irregularities. For real videos, by contrast, audio cues played a more important role, followed by visual appearance and prior knowledge. This may reflect the fact that authentic speech patterns and natural language flow are not easy for current generative models to mimic ~\cite{groh2024human}.

These insights have direct implications for designing media literacy interventions. Current programs often provide prescribing specific cues or providing guidelines~\cite{chesney2019deep, tahir2021seeing}, while assistive tools highlight AI-detected artefacts ~\cite{arrieta2020explainable, boyd2023value}. Our results suggest moving beyond "spot the glitch" approaches towards training that helps people recognise when their confidence may be misplaced. For example, visual appearance is useful for spotting surface-level artefacts but can be misleading when glitches are subtle or absent. Media literacy programs could therefore help users understand not only how to identify visual indicators of manipulation, but also when relying on certain visual cues may lead to false confidence. In addition, audio strategies (e.g., vocal tone, language coherence, sound quality) demonstrated to be more diagnostic for real videos than for deepfakes. Interventions could highlight this asymmetry, teaching users that while audio cues can confirm authenticity, they are less effective for spotting manipulation and may lead to calibration errors. Moreover, knowledge strategies (e.g., prior knowledge, external sources, intuition) were critical for both real and fake detection. However, reliance on intuition alone risks overconfidence. Interventions could emphasise fact-checking and cross-referencing external information as stronger complements to perceptual cues.

\subsection{Detection Challenges}
As generative AI advances, deepfakes may become perceptually indistinguishable from real videos. Our results already suggest this direction: multimodal strategies helped with real videos but offered little benefit for identifying deepfakes. 
Even with these limitations, whether humans can distinguish real from fake content is still important. Although AI detectors can support verification, they often lag behind new generative techniques and are not always accessible to the public. For example, commercial tools require payment, while open-source models demand technical expertise~\cite{wu2025understanding}. This highlights the continuing need to train people to recognise when their confidence may be misplaced and when additional verification is necessary. 

In scenarios involving political accountability, journalistic verification, or interpersonal trust, human judgment will remain essential. This is because deepfake disinformation can cause serious harm, such as damaging personal and organisational reputations, destabilising political processes, facilitating financial fraud, and enabling non-consensual pornography~\cite{vaccari2020deepfakes}. In these settings, human judgment acts as a first line of defence, especially when AI tools are unavailable or when rapid decisions are required. Furthermore, trained individuals may identify cues that are overlooked by AI models. In a Human-AI collaboration approach, integrating human evaluation with AI detectors may enhance the overall detection effectiveness and strengthen our capacity to counter deepfakes.

However, as deepfakes become more realistic, broader safeguards are required, including provenance or watermarking standards, platform-level disclosure of synthetic media, and regulations that penalise malicious deepfake creation and distribution~\cite{lyu2024deepfake, romero2025deepfake}. For example, recent regulations such as the UK’s Online Safety Act 2023, which places responsibility on platforms, and China’s Deep Synthesis Provisions, which require mandatory labelling and algorithmic review, illustrate how governments are beginning to address the risks posed by deepfakes~\cite{romero2025deepfake}.

\section{Conclusion}
This paper examined how people detect deepfake videos by analysing accuracy, confidence, and calibration across visual, audio, and knowledge-based strategies. We showed that while real videos benefited from multimodal integration of cues, deepfakes disrupted these strategies, leaving participants both being less accurate and more miscalibrated. By further breaking strategies into cues and mapping their co-occurrence networks, we identified which combinations supported or hindered detection and highlighted the central cues in supporting effective judgments. Together, these findings advance our understanding of human deepfake detection and suggest concrete directions for designing media literacy interventions that build on people's natural strategies.

This work has limitations. Our participant pool was restricted to young adults, which limits generalisability across age groups. The stimulus set included a limited number of deepfake videos and may not capture the full diversity of manipulations encountered in the wild. Future research can extend this work by including broader populations and more diverse video manipulations. Moreover, participants selected cues from a predefined checklist. While this structure enabled quantitative analysis, it may have prompted participants to report cues they might not use spontaneously. Future work could incorporate unprompted methods to capture more naturalistic strategies.

\bibliographystyle{ACM-Reference-Format}
\bibliography{sample-base}

\newpage
\appendix
\section*{Appendix A. Strategy Types and cues}
\begin{table}[ht]
\small
\begin{center}
\Description{Details about the relation between strategies and cues, as well as the description of cues.}
\begin{tabular}{|p{1.2cm}|p{1.3cm}|p{4.5cm}|}
\hline
\textbf{Strategy Type} & \textbf{Cue} & \textbf{Description} \\
\hline
\multirow{14}{*}{\parbox{1.2cm}{Visual}} 
    & \multirow{4}{*}{\parbox{1.3cm}{Appearance}} 
        & Facial features (e.g., uneven skin tone, facial symmetry, hairstyle, eyes not rendered well) \\
    & & Physical and behavioural characteristics (e.g., lip sync problems, abnormal body gestures, face swapping, stiffness/expressionless) \\
    \cline{2-3}
    & \multirow{6}{*}{\parbox{1.3cm}{Environment}} 
        & Characteristics of objects in video (e.g., oddly shaped artefacts, unnatural reflections, overlapping objects) \\
    & & Background and environmental details (e.g., scene settings, unusual backgrounds, issues with watermarks, logos or subtitles) \\
    & & Colour and lighting inconsistencies (e.g., discolouration, lighting inconsistencies, contrast problems) \\
    \cline{2-3}
    & \multirow{4}{*}{\parbox{1.3cm}{Production}} 
        & Video quality issues (e.g., video resolution, distortions, glitches, pixelation) \\
    & & Production issues (e.g., editing issues, camera angle/work issues, shakiness and jitter) \\
\hline
\multirow{8}{*}{\parbox{1.2cm}{Audio}} 
    & \multirow{2}{*}{\parbox{1.3cm}{Vocal}} & Vocal features (e.g., shift in pitch and tone, naturalness of voice, lack of emotions) \\
    \cline{2-3}
    & \multirow{2}{*}{\parbox{1.3cm}{Language}} & Intelligibility and language (e.g., mismatched in language used, fluency, pronunciation, loss of intelligibility) \\
    \cline{2-3}
    & \multirow{4}{*}{\parbox{1.3cm}{Audio quality}} 
        & Sound quality issues (e.g., clarity problems, uneven volume, distortion, echoes, glitches) \\
    & & Background sound issues (e.g., background reverberation/echoes, overall noise, mechanical noises) \\
\hline
\multirow{13}{*}{\parbox{1.2cm}{Knowledge}} 
    & \multirow{4}{*}{\parbox{1.3cm}{Information source}} 
        & Online tools (e.g., using search engines like Google or social media platforms like Facebook and YouTube) \\
    & & Use of multiple sources \\
    \cline{2-3}
    & \multirow{7}{*}{\parbox{1.3cm}{Prior knowledge}} 
        & Knowledge of the person (e.g., prior knowledge of the people in the video) \\
    & & Knowledge of video content (e.g., familiarity with the event in the video) \\
    & & Knowledge of the context surrounding the video's content \\
    \cline{2-3}
    & \multirow{2}{*}{\parbox{1.3cm}{Intuition}} & Intuition and emotions (e.g., based on your own instinct, opinions or emotions) \\
\hline
\end{tabular}
\caption{Strategy Types and Corresponding cues}
\label{tab:strategy-subcues}
\end{center}
\end{table}

\newpage
\section*{Appendix B. Video Information}
\begin{table}[ht]
\begin{center}
\Description{Details about the information of videos used in the study}
\begin{tabular}{|p{1.1cm}|p{1.7cm}|p{3.8cm}|p{0.9cm}|}
\hline
\textbf{Video Type} & \textbf{Topic} & \textbf{Description} & \textbf{Length} \\
\hline

\multirow{10}{*}{Deepfake}
    & \multirow{4}{*}{Entertainment}
        & Mark Zuckerberg says he controls billions of peoples' confidential data and thus owns their future. & 00:18 \\
    & & Kim Kardashian tells how she likes making money by manipulating her fans. & 00:22 \\
    & & Tom Cruise's daily life. & 00:41 \\
    & & A clip from \textit{The Shining}. & 00:51 \\
    \cline{2-4}

    & \multirow{3}{*}{Politics}
        & Manoj Tiwari criticises an opposing political party and encourages people to vote for his party. & 00:48 \\
    & & Jeremy Corbyn supports Boris Johnson as Prime Minister. & 00:33 \\
    & & A speech announcing that the Apollo 11 mission had failed. & 01:03 \\
    \cline{2-4}

    & \multirow{2}{*}{Educational}
        & Obama reminds people to be more alert to fake news. & 00:27 \\
    & & Morgan Freeman asks: "Is seeing believing?" & 00:28 \\
    \cline{2-4}

    & Sports
        & Jose Mourinho comments on soccer. & 01:27 \\
\hline

\multirow{10}{*}{Real}
    & \multirow{4}{*}{Entertainment}
        & Mark Zuckerberg says he can ascertain people's online behaviours. & 00:29 \\
    & & Kim Kardashian claims she cheated on an exam. & 00:15 \\
    & & Tom Holland takes a break from social media. & 00:51 \\
    & & Another clip from \textit{The Shining}. & 00:50 \\
    \cline{2-4}

    & \multirow{3}{*}{Politics}
        & Biden criticises MAGA Republicans. & 00:51 \\
    & & Trump urges his supporters to calm down following congressional attacks. & 00:51 \\
    & & President Uhuru mourns former Kenyan President Mwai Kibaki. & 00:42 \\
    \cline{2-4}

    & \multirow{2}{*}{Educational}
        & Hillary Clinton discusses the dangers of fake news. & 00:56 \\
    & & Ellen warns people about fake news. & 00:29 \\
    \cline{2-4}

    & Sports
        & Jose Mourinho discusses Sir Alex Ferguson's reaction after Porto's Champions League win over Manchester United. & 01:17 \\
\hline

\end{tabular}
\caption{Videos Used in the Study}
\label{tab:video-descriptions}
\end{center}
\end{table}

\begin{figure}[t]
  \centering
  \Description{A screenshot of a deepfake video of Jeremy Corbyn}
  \includegraphics[width=\linewidth]{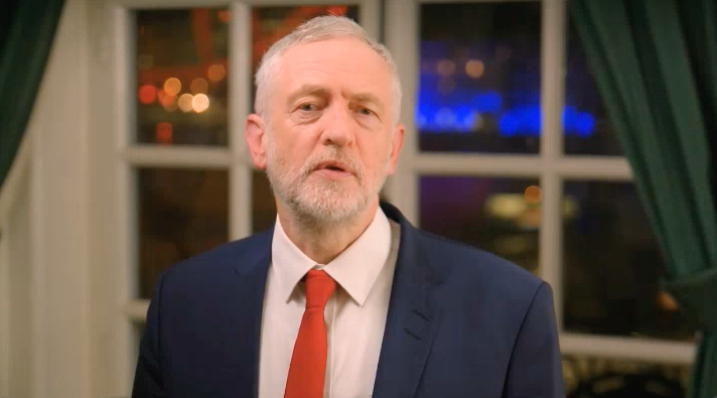}
  \caption{Deepfake video of Jeremy Corbyn}
  \label{fig:jeremy}
\end{figure}

\begin{figure}[t]
  \centering
  \Description{A screenshot of a deepfake video of Mark Zuckerberg}
  \includegraphics[width=\linewidth]{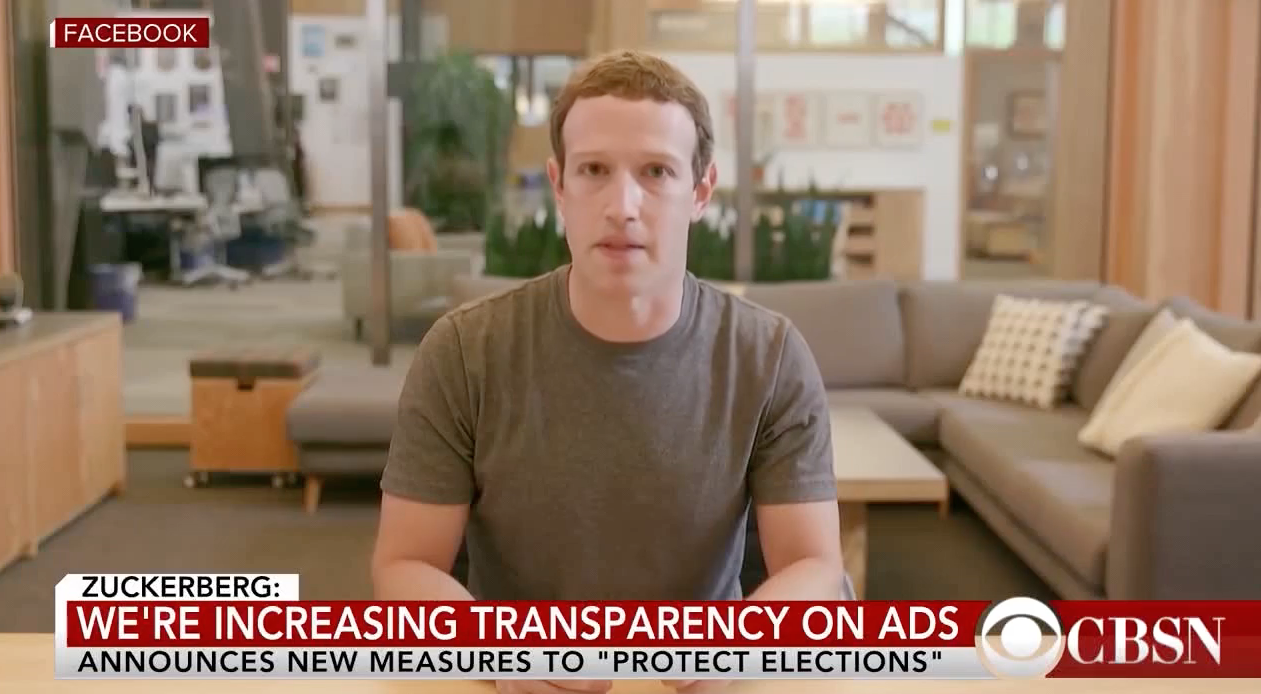}
  \caption{Deepfake video of Mark Zuckerberg}
  \label{fig:mark}
\end{figure}

\begin{figure}[t]
  \centering
  \Description{A screenshot of a real video of Hillary Clinton}
  \includegraphics[width=\linewidth]{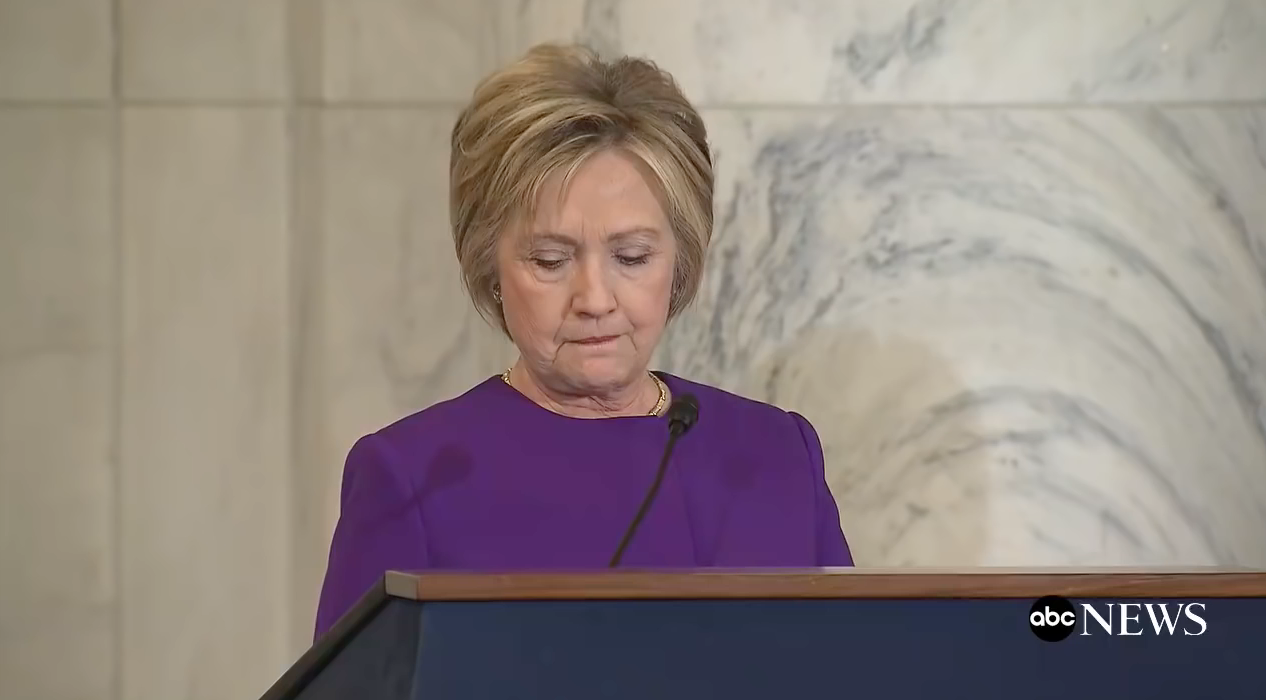}
  \caption{Real video of Hillary Clinton}
  \label{fig:hillary}
\end{figure}

\begin{figure}[H]
  \centering
  \Description{A screenshot of a real video of Trump}
  \includegraphics[width=\linewidth]{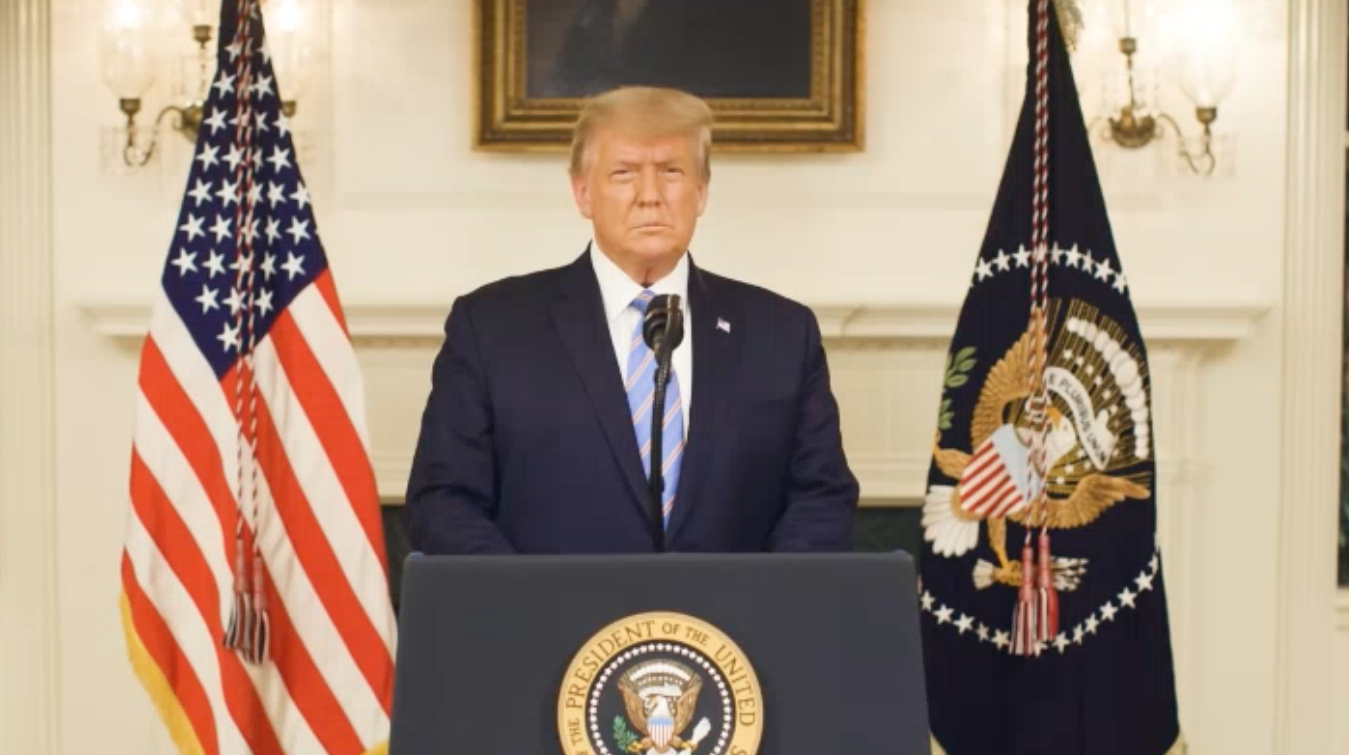}
  \caption{Real video of Trump}
  \label{fig:trump}
\end{figure}

\section*{Appendix C. Questionnaire Items}
\noindent\textbf{Repeated for each of the 4 videos}

\begin{enumerate}
    \item \textbf{Is the video real or fake?}
    \begin{itemize}
        \item Real
        \item Fake
    \end{itemize}

    \item \textbf{Confidence Rating} \\
    What is your level of confidence in the above selection?  
    Please indicate your confidence on a scale of 1 to 5.
    \begin{itemize}
        \item 1
        \item 2
        \item 3
        \item 4
        \item 5
    \end{itemize}

    \item \textbf{Visual Characteristics} \\
    Using what you \textit{saw} in the video, which visual cues helped you decide whether the video was real or fake?
    \begin{itemize}
        \item Facial features (e.g., uneven skin tone, facial symmetry, hairstyle, eyes not rendered well)
        \item Physical and behavioural characteristics (e.g., lip sync problems, abnormal body gestures, face swapping, stiffness/expressionlessness)
        \item Characteristics of objects in the video (e.g., oddly shaped artefacts, unnatural reflections, overlapping objects)
        \item Background and environmental details (e.g., scene settings, unusual backgrounds, issues with watermarks, logos or subtitles)
        \item Video quality issues (e.g., resolution, distortions, glitches, pixelation)
        \item Production issues (e.g., editing issues, camera angle issues, shakiness or jitter)
        \item Colour and lighting inconsistencies (e.g., discolouration, inconsistent lighting, contrast problems)
        \item Any other attributes (please specify)
    \end{itemize}

    \item \textbf{Audio Characteristics} \\
    Using what you \textit{heard} in the video, which audio cues helped you decide whether the video was real or fake?
    \begin{itemize}
        \item Vocal features (e.g., shift in pitch or tone, naturalness of voice, lack of emotion)
        \item Intelligibility and language (e.g., mismatched language, fluency, pronunciation, loss of intelligibility)
        \item Sound quality issues (e.g., clarity problems, uneven volume, distortion, echoes, glitches)
        \item Background sound issues (e.g., reverberation, mechanical noise, overall background noise)
        \item Any other attributes (please specify)
    \end{itemize}

    \item \textbf{Knowledge-Based Characteristics} \\
    Using your own knowledge or external information, which cues helped you decide whether the video was real or fake?
    \begin{itemize}
        \item Online tools (e.g., searching via Google, Facebook, YouTube)
        \item Communication with others (e.g., checking with family or friends offline/online)
        \item Use of multiple sources 
        \item Knowledge of the person (e.g., prior familiarity with the individual in the video)
        \item Knowledge of the video content (e.g., familiarity with the event shown)
        \item Knowledge of the surrounding context (e.g., broader context, consistency of meaning, political affiliations)
        \item Intuition and emotions (e.g., instinct, personal opinions, emotional impression)
        \item Any other attributes (please specify)
    \end{itemize}
\end{enumerate}

\section*{Appendix D. Study Interface}
\begin{figure}[ht]
  \centering
  \Description{A screenshot of the study interface}
  \includegraphics[width=\linewidth]{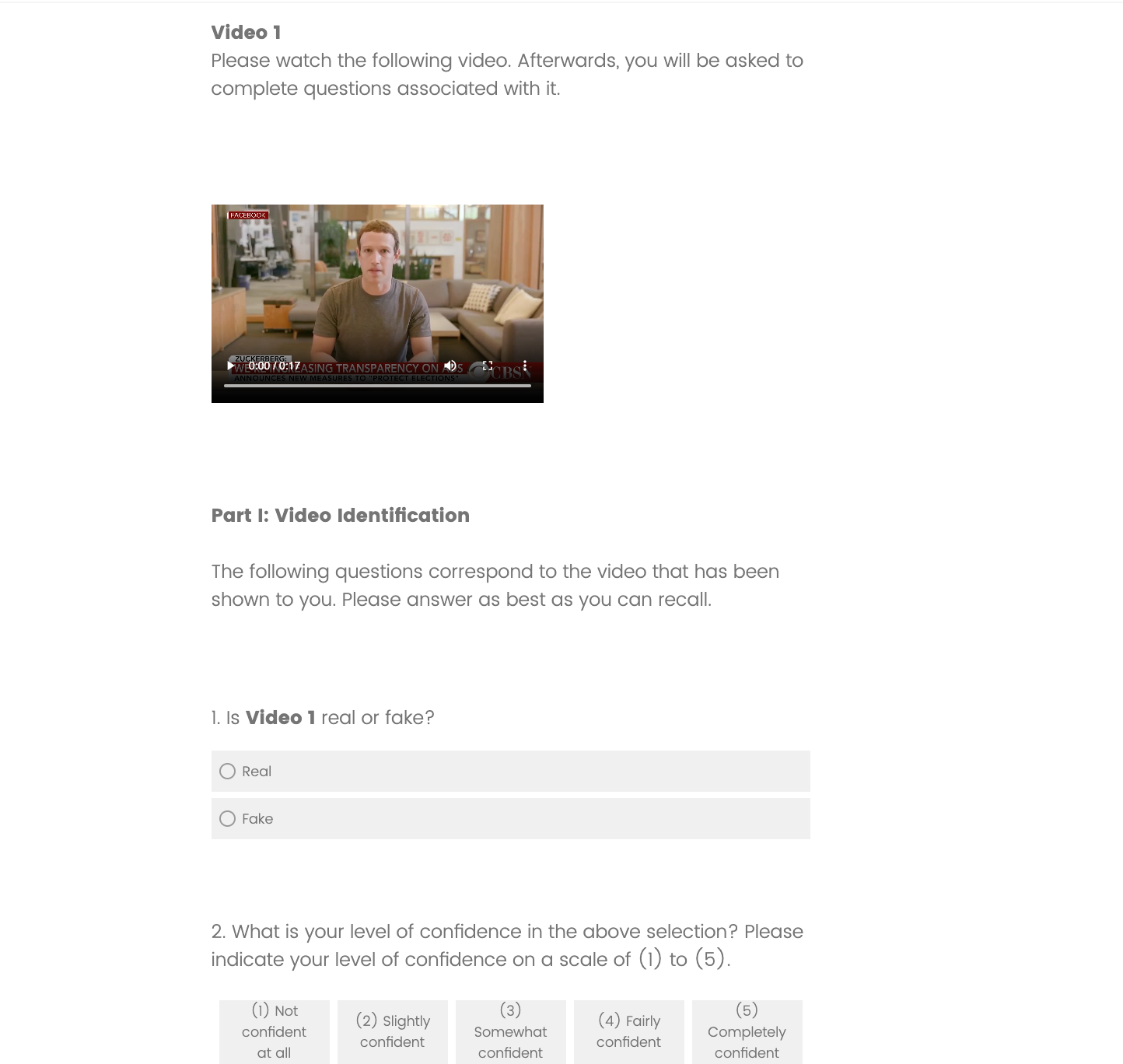}
  \caption{Study Interface}
  \label{fig:interface}
\end{figure}

\appendix
\end{document}